\documentclass[aps,twocolumn,prd,nofootinbib,showpacs]{revtex4}
\usepackage{amsmath}
\usepackage{amssymb}
\usepackage[aps]{mrv}
\usepackage{graphicx}
\usepackage{hyperref}
\usepackage{color}

\makeatletter
\gdef\@fpheader{}
\g@addto@macro\bfseries{\boldmath}
\makeatother

\newcommand{\thetai}{\boldmathsymbol{\theta}_{\uinf}}
\newcommand{\thetar}{\boldmathsymbol{\theta}_{\ureh}}

\newcommand{\si}{\mathrm{SI}}

\newcommand{\li}{\mathrm{LI}}
\newcommand{\di}{\mathrm{DI}}
\newcommand{\sfi}{\mathrm{SFI}}
\newcommand{\pli}{\mathrm{PLI}}
\newcommand{\lmi}{\mathrm{LMI}}
\newcommand{\kmiii}{\mathrm{KMIII}}
\newcommand{\cwi}{\mathrm{CWI}}
\newcommand{\osti}{\mathrm{OSTI}}

\newcommand{\alphali}{\alpha_{\eisss{\li}}}
\newcommand{\alphalimax}{\alphali^{\max}}
\newcommand{\alphalimin}{\alphali^{\min}}

\newcommand{\bardeen}{\Phi_{\eisss{\mathrm{B}}}}
\newcommand{\pnad}{P_{\mathrm{nad}}}
\newcommand{\cS}{c_{\eisss{\uS}}}

\newcommand{\calLmax}{\calL_{\max}}

\hypersetup{
    bookmarks=true,         
    unicode=false,          
    pdftoolbar=true,        
    pdfmenubar=true,        
    pdffitwindow=false,     
    pdfstartview={FitH},    
    pdftitle={My title},    
    pdfauthor={Author},     
    pdfsubject={Subject},   
    pdfcreator={Creator},   
    pdfproducer={Producer}, 
    pdfkeywords={keyword1} {key2} {key3}, 
    pdfnewwindow=true,      
    colorlinks=true,       
    linkcolor=red,          
    citecolor=cyan,        
    filecolor=magenta,      
    urlcolor=blue,           
    linktocpage=true
}

\begin{document}

\title{Shortcomings of New Parametrizations of Inflation}

\author{J\'er\^ome Martin} \email{jmartin@iap.fr}
\affiliation{Institut d'Astrophysique de Paris, UMR 7095-CNRS,
Universit\'e Pierre et Marie Curie, 98 bis boulevard Arago, 75014
Paris, France}

\author{Christophe Ringeval} \email{christophe.ringeval@uclouvain.be}
\affiliation{Centre for Cosmology, Particle Physics and Phenomenology,
  Institute of Mathematics and Physics, Louvain University, 2 Chemin
  du Cyclotron, 1348 Louvain-la-Neuve, Belgium}

\author{Vincent Vennin} \email{vincent.vennin@port.ac.uk}
\affiliation{Institute of Cosmology \& Gravitation, University of
  Portsmouth, Dennis Sciama Building, Burnaby Road, Portsmouth, PO1
  3FX, United Kingdom}

\date{\today}

\begin{abstract}
  In the hope of avoiding model dependence of the cosmological
  observables, phenomenological parametrizations of Cosmic Inflation
  have recently been proposed. Typically, they are expressed in terms
  of two parameters associated with an expansion of the inflationary
  quantities matching the belief that inflation is characterized by
  two numbers only, the tensor-to-scalar ratio and the scalar spectral
  index. We give different arguments and examples showing that these
  new approaches are either not generic or insufficient to make
  predictions at the accuracy level needed by the cosmological
  data. We conclude that disconnecting inflation from high energy
  physics and gravity might not be the most promising way to learn
  about the physics of the early Universe.
\end{abstract}

\maketitle

\section{Introduction}
\label{sec:intro}

With the advent of precision cosmology, it is now possible to
observationally probe the early Universe and its front-runner
paradigm, Cosmic Inflation~\cite{Starobinsky:1979ty,
  Starobinsky:1980te, Guth:1980zm, Linde:1981mu, Albrecht:1982wi,
  Linde:1983gd, Mukhanov:1981xt, Mukhanov:1982nu, Starobinsky:1982ee,
  Guth:1982ec, Hawking:1982cz, Bardeen:1983qw}. When the mechanism of
inflation was discovered, only a few
models~\cite{Starobinsky:1980te,Albrecht:1982wi,Linde:1983gd}, making
simple predictions, were proposed. However, over time, many more
scenarios, often complex, were devised. This has resulted in a
situation where literally hundreds of inflationary models are a priori
possible. This should not come as a surprise given that, in order to
build an inflationary model, one has to extrapolate high energy
physics, or gravity, by many orders of magnitude, in a regime where
nothing is experimentally known. In some sense, the profusion of
proposed models is due to our lack of knowledge of physics beyond the
electroweak scale and not to a lack of predictability of inflation.

However, with the recent release of the Planck 2013 \& 2015
data~\cite{Ade:2013ktc, Adam:2015rua, Ade:2015xua, Ade:2015lrj}, the
Augean stables have started to be cleaned up. Indeed, models of
inflation generating non-negligible isocurvature perturbations, large
non-Gaussianities and/or significant features in the power spectrum
are, for the moment, disfavored by observations.  Single-field
slow-roll models of inflation with a minimal kinetic term therefore
appear to be preferred~\cite{Martin:2010hh, Easther:2011yq,
  Martin:2013tda, Martin:2013nzq, Price:2015qqb, Martin:2015dha}, even
if a large number of other scenarios still remain compatible with the
data~\cite{Chen:2012ja, Vennin:2015vfa, Vennin:2015egh, Chen:2016vvw}.

An alternative approach to systematic model comparison consists in
considering model independent parametrizations of inflation. Such
parametrizations aim at embracing all models at once while avoiding
difficult questions related to specifying a potential $V(\phi)$, as
for instance discussing the physical values of its parameters,
possible quantum corrections or interaction of the inflaton field with
other sectors. Such a proposal was first implemented within
the slow-roll formalism~\cite{Mukhanov:1985rz, Mukhanov:1988jd,
  Stewart:1993bc, Gong:2001he, Schwarz:2001vv, Leach:2002ar}. It has
been successfully applied to models with non-minimal kinetic
terms~\cite{Kinney:2007ag, Tzirakis:2008qy, Lorenz:2008et,
  Agarwal:2008ah, Martin:2013uma, Jimenez:2013xwa}, multifield
inflation and modified gravity~\cite{Nakamura:1996da, Easther:2005nh,
  DiMarco:2005nq, Battefeld:2006sz, Chiba:2008rp, DeFelice:2011bh}
while being used for non-Gaussianities~\cite{Chen:2006nt,
  Yokoyama:2007uu, Ichikawa:2008iq, Langlois:2008qf, Chen:2010xka} as
well. Classes of inflationary models could also be devised owing to slow
roll, as for instance the Schwarz and Terrero-Escalante (STE)
classification~\cite{Schwarz:2004tz} where only one of the three
classes survived the Planck measurements~\cite{Martin:2013nzq}. If the
microphysics is considered instead, the effective theory of
inflation~\cite{Creminelli:2004yq, Cheung:2007st} can also be a way
to parametrize deviations from the simplest physical setups.

These parametrizations yield a vast range of observable predictions,
precisely because they are intended to be model independent and
designed to describe many possible scenarios. However knowing a
preferred range for the tensor-to-scalar ratio $r$ would greatly help
the design of future missions aiming at measuring the $B$-polarization
of the Cosmic Microwave Background (CMB) radiation. Similarly, the
amount of non-Gaussianities expected within various classes of
inflationary models is valuable information for future galactic
surveys such as Euclid~\cite{Amendola:2016saw}.

For these reasons and despite the existence of the slow-roll
formalism, new ``simple'' parametrizations have recently been
proposed, that aim at narrowing down inflationary
predictions. Moreover, it has been suggested that, at the
observational level, inflation can be reduced to two numbers only (the
scalar power spectrum spectral index and the tensor-to-scalar ratio),
which was argued to further motivate the introduction of these new
frameworks. These new parametrizations include, among others, the
truncated horizon-flow formalism~\cite{Hoffman:2000ue, Kinney:2002qn,
  Ramirez:2005cy, Chongchitnan:2005pf}, the ``universality
classes''~\cite{Huang:2007qz, Roest:2013fha, Garcia-Bellido:2014gna,
  Binetruy:2014zya, Huang:2015cke}, and designing a simple
hydrodynamical description of inflation~\cite{Mukhanov:2013tua,
  Creminelli:2014nqa}.

In this short article, we investigate whether these new approaches can
be further used to constrain the physics of the early Universe (for
issues with the truncated horizon-flow approach, see
Refs.~\cite{Liddle:2003py, Vennin:2014xta, Coone:2015fha}). The paper
is organized as follows. In \Sec{sec:nottrueapprox}, we explain why
expanding inflationary observables in the so-called ``large $N$
limit'' ($N$ being the number of e-folds) is not always consistent. We
also show that the number of ``universality classes'' becomes large
beyond the leading order where they thus provide a more complex
classification. In \Sec{sec:notsufficient}, we show that the large $N$
limit gives insufficiently accurate predictions for the spectral index
$\nS$ and the tensor-to-scalar ratio $r$. With respect to the Planck
2015 confidence intervals, these inaccuracies range from one to two
sigma or more, depending on the underlying inflationary scenario. In
\Sec{sec:reheat}, it is shown that this approach does not allow one to
consistently incorporate reheating, nor to derive constraints on its
expansion history. \Sec{sec:eosi} is dedicated to the alternative
parametrization of inflation in which one specifies the equation of
state parameter $w(N)$ as a function of the number of
e-folds~\cite{Mukhanov:2013tua}. Such a parametrization is shown to be
free of the above-mentioned issues for the simple reason that, at the
background level, it ends up being equivalent to choosing a specific
potential for a single scalar field. At the perturbative level, it is
either incomplete because the speed of sound and the non-adiabatic
pressure have to be specified (see also Ref.~\cite{Chen:2013kta}), or
implicitly equivalent to a perturbed single scalar field. In
\Sec{sec:stat}, we stress the fact that all these alternative
approaches, independently of their internal consistencies, are not
well suited to perform Bayesian statistical analysis of the
cosmological data. Finally, in the conclusion, we argue that these
frameworks do not allow one to connect inflation and high energy
physics (modified gravity included).
\section{Not universal}
\label{sec:nottrueapprox}
\subsection{General definitions}
\label{sec:genedef}
In the standard formulation, a single-field slow-roll model of
inflation is specified by a potential $V(\phi)$. Then, the behavior of
the system is completely controlled by the Friedmann-Lema\^itre and
Klein-Gordon equations. In general, these equations cannot be solved
analytically and one has to use either an exact numerical
integration~\cite{Salopek:1988qh, Grivell:1999wc, Adams:2001vc,
  Tsujikawa:2002qx, Parkinson:2004yx, Makarov:2005uh, Ringeval:2007am}
or an approximation scheme. Given that, during inflation, the Hubble
parameter $H$ is almost constant, one can define an analytical
expansion in terms of small parameters that are the successive
derivatives of $H$. These are called ``Hubble-flow'' parameters and
are given by~\cite{Schwarz:2001vv}
\begin{equation}
\eps{n+1}=\frac{\dd \ln \vert \eps{n}\vert}{\dd N}, \quad 
n \ge 0,
\end{equation}
where $N=\ln (a/\aini)$ is the number of e-folds and $\eps{0}\equiv
\Hini/H$. Using these functions, one can then perturbatively calculate
the power spectra of scalar and tensor modes. At leading order, the
expressions of the scalar spectral index, tensor-to-scalar ratio and
scalar running are given by~\cite{Stewart:1993bc, Gong:2001he,
  Schwarz:2001vv, Leach:2002ar, Jimenez:2013xwa, Martin:2013uma}
\begin{eqnarray}
\label{eq:ns}
\nS & = & 1 - 2\eps{1*} -\eps{2*} +\order{\epsilon^2}, \\
\label{eq:r}
r  & = & 16 \eps{1*} +\order{\epsilon^2}, \\
\label{eq:running}
\alphaS & = & \order{\epsilon^2}.
\end{eqnarray} 
These formulas are evaluated at the field value $\phi_*$
where the pivot scale at which these quantities are defined exits the
Hubble radius during inflation. It is expressed in terms of $\Delta N_*\equiv
\Nend - \Nstar$ as
\begin{equation}
\label{eq:traj:SR:LO}
\Delta N_* = -\frac{1}{\Mp^2}\int_{\phi_*}^{\phi_\uend}
\frac{V}{V^\prime} \, \ud\phi,
\end{equation}
where $\phi_\uend$ satisfies $\eps{1}(\phi_\uend)=1$ and denotes the
value of $\phi$ at the end of inflation. Here, primes denote
differentiation with respect to $\phi$. 

One can also introduce ``slow-roll'' parameters, noted
$\epsV{n}$ in the following, and defined directly from the inflaton
potential and its derivatives, namely~\cite{Liddle:1994dx}
\begin{equation}
\epsV{1}= \frac{\Mp^2}{2}\left(\frac{V^\prime}{V}\right)^2 ,\quad
\epsV{2} = 2\Mp^2 \left[\left(\frac{V^\prime}{V}\right)^2 
- \frac{V^{\prime\prime}}{V}\right],
\label{eq:epsV:V}
\end{equation}
the hierarchy, as for the Hubble-flow functions, being also infinite
with, for instance, $\epsV{2} \epsV{3}=- \Mp^2 \epsV{2}' V'/V$,
$\epsV{3}\epsV{4}=- \Mp^2 \epsV{3}' V'/V$ and so on. At leading order,
one can show that 
\begin{equation}
\eps{1}=\epsV{1}, \quad \eps{2}=\epsV{2}.
\label{eq:epsV:eps:LO}
\end{equation}

In practice, for a given potential $V(\phi)$, one first
calculates the functions $\epsV{1}(\phi)$ and $\epsV{2}(\phi)$. As just
explained, this directly leads to the Hubble flow functions $\eps{1}$
and $\eps{2}$ through \Eq{eq:epsV:eps:LO}. The end of inflation can 
be determined from the
condition $\eps{1}(\phi_\uend)=\epsV{1}(\phi_\uend)=1$. Notice that, a
priori, using $\eps{1}=\epsV{1}$ at the end of inflation is not
justified since, by definition, slow roll is violated in this
regime. But, in practice, this leads to a small error that can be
neglected. Finally, \Eq{eq:traj:SR:LO} allows us to derive
$\phi_*=\phi(\Delta N_*)$ and, putting everything together, one
arrives at
\begin{equation}
\eps{n*}=\epsV{n*}=
\epsV{n}(\phi_*)=\epsV{n}\left[\phi(\Delta
  N_*)\right].
\end{equation}
As a consequence, choosing a value for $\Delta N_*$ (which depends on the
reheating epoch and the post-inflationary history of the universe, see
\Sec{sec:reheat}) leads to a definite value for $\eps{n*}$ and,
therefore, to a prediction of the model since $\nS$ and $r$ are now
explicitly known.

\par

At the time of precision cosmology, it is in fact important to go to
next-to-leading order. This is highly nontrivial since this causes
new problems. For instance, the calculation of the power spectra
becomes much more involved because the standard method, based on the
Bessel function method, is no longer available. Fortunately, there
exist other methods, for instance based on the
Wentzel-Kramers-Brillouin approximation (or its extension such as the
uniform approximation), which allows one to go beyond the leading
order. This leads to the following expressions~\cite{Gong:2001he,
  Schwarz:2001vv, Jimenez:2013xwa, Martin:2013uma}
\begin{align}
  \label{eq:ns2}
\nS & =  1 - 2\eps{1*} -\eps{2*}  - (3+2C)
\eps{1*} \eps{2*} 
\nonumber \\ 
& - 2 \eps{1*}^2 - C \eps{2*} \eps{3*}+\order{\epsilon^3}, \\
\label{eq:r2}
r  & =  16 \eps{1*} \left (1 + C \eps{2*} \right)+\order{\epsilon^3}, \\
\label{eq:running2}
\alphaS & =  -2 \eps{1*} \eps{2*} - \eps{2*} \eps{3*}+\order{\epsilon^3},
\end{align}
where $C \equiv \gamma + \ln 2 -2$, $\gamma$ being the Euler
constant. Another modification that arises at next-to-leading order is
that the functions $\eps{n}$ and $\epsV{n}$ no longer
coincide. Indeed, from the slow-roll parameters~(\ref{eq:epsV:V})
calculated by means of the potential, the Hubble-flow functions at
second order are given by~\cite{Liddle:1994dx, Vennin:2014xta}
\begin{equation}
\begin{aligned}
\eps{1} & = \epsV{1} \left(1 - \dfrac{\epsV{2}}{3} \right) 
+ \order{\epsV{}^3},\\
\eps{2} & = \epsV{2} \left(1 - \dfrac{\epsV{2}}{6} -
\dfrac{\epsV{3}}{3}\right)  + \order{\epsV{}^3}, \\ 
\eps{3} & = \epsV{3} \left(1 - \dfrac{\epsV{2}}{3} -
\dfrac{\epsV{4}}{3} \right) + \order{\epsV{}^3}.
\end{aligned}
\label{eq:epsVtoH}
\end{equation}
A priori, this also means that the determination of the end point of
inflation is modified since, at this order, we no longer have
$\eps{1}=\epsV{1}$. However, we have already seen that the slow-roll
approximation is anyway violated for $\eps{1}=1$ and, therefore,
adding a correction in this regime cannot be trusted. For this reason,
one still uses the condition $\epsV{1}=1$, as well as
\Eq{eq:traj:SR:LO}, to determine the end of inflation and the
trajectory, respectively. The error induced on $\Delta \Nstar$ ends up
being small, of a few e-folds at most~\cite{Salopek:1988qh, Grivell:1999wc, Adams:2001vc,
  Tsujikawa:2002qx, Parkinson:2004yx, Makarov:2005uh,
  Ringeval:2007am}, the reason being that when slow roll is violated
inflation cannot be sustained for many e-folds.

Following the same logics as explained before, one can finally find
the function $\eps{n*}=\eps{n}(\Delta N_*)$, this time at
next-to-leading order. In this way, one can obtain more accurate
predictions if needed.
\subsection{One universality class}
\label{sec:oneuc}
Recently, various works have tried to parametrize inflation by a first
order expansion of $\epsstar{1}$ in the small number $1/\Delta
\Nstar\ll 1$. Originally, it was postulated that most interesting
inflationary scenarios should lead to~\cite{Roest:2013fha}
\begin{align}
\epsstar{1}=\frac{\beta}{\left(\Delta \Nstar\right)^\alpha}+\cdots ,
\label{eq:expeps1}
\end{align}
the higher order terms being assumed to be negligible. The motivation
in doing so is the remark that, assuming $\Delta \Nstar=\order{10^2}$,
the deviations expected from scale invariance for the simplest case
$\alpha=1$ are of the order $10^{-2}$, which is, up to a factor of a
few, the current measurement of the spectral index $\nS-1$. 

In fact, it is easy to find models for which \Eq{eq:expeps1} is not true. For
instance, Kh\"aler Moduli Inflation II ($\kmiii$),
\begin{equation}
V(\phi) \propto 1 - \bar{\alpha} \left(\dfrac{\phi}{\Mp}\right)^{4/3} \exp\left[-\bar{\beta}
  \left(\dfrac{\phi}{\Mp}\right)^{4/3}\right],
\end{equation}
where $\bar{\alpha}$ and $\bar{\beta}$ are two model parameters, one of
the best models according to the Planck data (this model belongs to the
``plateau inflation'' category), leads to~\cite{Martin:2013tda}
\begin{equation}
\begin{aligned}
  \epsVstar{1} &= \frac{\ln ^{5/2}\left(16\bar{\alpha}
\sqrt{\dfrac{9\bar{\beta} ^{1/2}}{8}}  \Delta \Nstar\right)}
{324 \bar{\beta}^{3/2}{\Delta \Nstar}^2}
+\order{\frac{1}{{\Delta\Nstar}^3}}, \\
\epsVstar{2} &=\frac{2}{\Delta \Nstar}
+\order{\frac{1}{{\Delta\Nstar}^2}}.
\end{aligned}
\end{equation}
Let us stress that this model is not a contrived scenario designed to
artificially produce a dependence different from the one of
Eq.~(\ref{eq:expeps1}). It is a string-inspired model that fits the
data very well~\cite{Martin:2013tda}, and is one example among others
for which the first Hubble flow function does not scale as an inverse
power law of $\Delta N_*$.
\subsection{Several classes}
\label{subsec:severaluc}
The number of ``universality classes'' was then extended in
Ref.~\cite{Garcia-Bellido:2014gna}, in which power-law
terms~(\ref{eq:expeps1}) belong to the ``perturbative'' category,
purely exponential terms are ``non-perturbative'' and logarithm
functionals such as $\kmiii$ belong to the ``logarithmic'' class. This
shows that in practice, to design a complete set of universality
classes, one has to study all inflationary models, as done in
Refs.~\cite{Martin:2013tda, Martin:2013nzq}, compute their predictions
and, then, attempt to organize them into ``universality classes''.

As a consequence, universality classes do not dispense one with a
systematic study of theoretically motivated inflationary models. In
this sense, they do not seem more generic than the standard approach
but should rather be seen as a way to classify models, similar to the
STE classification~\cite{Schwarz:2004tz} for instance. In particular,
if new models are proposed in the future, the introduction of
additional universality classes may be necessary.

\subsection{Even more classes at next-to-leading order}
\label{sec:noeqClassNLO}
A leading order term of the form $\propto 1/(\Delta \Nstar)^{\alpha}$
does not guarantee that the next-to-leading order terms (the
importance of which will be demonstrated in \Sec{sec:notsufficient})
are similar and the expansion simple.  In this section, we show that a
model can be ``perturbative'' at leading order while being
``logarithmic'' at next-to-leading order for instance.  In principle,
this requires introducing a new classification at next-to-leading
order and further extending the number of classes that are necessary to
describe all situations.

Let us consider one of the simplest and currently favored models of
inflation, namely, the Starobinsky model~\cite{Starobinsky:1980te}
$\si$, for which the potential is given by
\begin{equation}
 V(\phi)\propto \left(1-e^{-\sqrt{2/3}\phi/\Mp} \right)^2.
\end{equation}
Up to the overall normalization, this model has no free parameter.
Jumping straight to the result, one obtains

\begin{widetext}
\begin{equation}
\begin{aligned}
\nS & = 1 - \dfrac{2}{\Delta \Nstar} + \dfrac{1}{(\Delta \Nstar)^2}
\left[ -\dfrac{5}{3} + \sqrt{3} - 2 C - \dfrac{3}{2} \ln \left(1+
  \dfrac{2}{\sqrt{3}} \right) + \dfrac{3}{2} \ln \left(\dfrac{4}{3}
  \Delta \Nstar \right) \right] + \order{\dfrac{1}{{\Delta \Nstar}^3}},\\
r & = \dfrac{12}{(\Delta \Nstar)^2} - \dfrac{2}{(\Delta \Nstar)^3}
\left[4 + 6 \sqrt{3} - 12 C - 9\ln\left(1+\dfrac{2}{\sqrt{3}} \right)
  + 9 \ln\left(\dfrac{4}{3} \Delta \Nstar \right) \right] +
\order{\dfrac{1}{{\Delta \Nstar}^4}},\\
\alphaS & = - \dfrac{2}{(\Delta \Nstar)^2} + \dfrac{1}{(\Delta
  \Nstar)^3} \left[-\dfrac{25}{6} + 2 \sqrt{3} - 3 \ln \left(1+
  \dfrac{2}{\sqrt{3}} \right) + 3 \ln \left(\dfrac{4}{3}
  \Delta \Nstar \right) \right] + \order{\dfrac{1}{{\Delta \Nstar}^4}}.
\end{aligned}
\label{eq:sipl}
\end{equation}
\end{widetext}
Although the first terms of the series are inverse power laws of
$\Delta \Nstar$, the higher order terms are not just given by higher
inverse power laws but also contain logarithms of $\Delta
\Nstar$. 
Let us also notice that
these expressions extend the ones of \Ref{Roest:2013fha}~[see
  Eq.~(32)].\footnote{It is interesting to explain how this was obtained. Defining
$x=\phi/\Mp$, at leading order, the $\si$ slow-roll trajectory reads
\begin{equation}
\xstar=\sqrt{\frac{3}{2}}\left[-\fstar
-\Lambert{-1}\left(-\ee^{-\fstar}\right)\right],
\end{equation}
where $\Lambert{-1}$ is the ``$-1$'' branch of the Lambert function and
\begin{equation}
\fstar \equiv \frac43 \Delta \Nstar -
\sqrt{\dfrac{2}{3}} \xend+\ee^{\sqrt{2/3}\xend},
\end{equation}
with $\xend\equiv \sqrt{3/2}\ln(1+2/\sqrt{3})$. Plugging this
trajectory into \Eq{eq:epsV:V}, and expanding it at
next-to-leading order, one obtains
\begin{eqnarray}
\label{eq:epsv1si}
\epsV{1*} &=\dfrac{3}{4\Delta N_*^2}
-\dfrac{9}{8\Delta N_*^3}
\left[\dfrac{2}{\sqrt{3}}-\ln \left(1+\dfrac{2}{\sqrt{3}}\right)
\right.
\nonumber \\ & \left.
+\ln \left(\dfrac43 \Delta N_*\right)\right]
+\order{\dfrac{1}{{\Delta\Nstar}^4}}.
\end{eqnarray}
The formula for $r$ found in \Ref{Roest:2013fha}, namely, 
\begin{equation}
r_{\text{\cite{Roest:2013fha}}}=\dfrac{12}{\Delta N_*^2}
-\dfrac{18}{\Delta N_*^3}
\ln \left(\Delta N_*\right),
\label{eq:sfi:r:Roest}
\end{equation}
corresponds to taking the expression~(\ref{eq:epsv1si}) and using it
in \Eq{eq:r}. \Eq{eq:sfi:r:Roest} is valid at order $\ln(\Delta\Nstar)/\Delta\Nstar^3$ but 
not at next-to-leading order in slow roll where higher 
$1/\Delta\Nstar^3$ terms appear. 
In particular, this amounts to ignoring the numerical
factor $2/\sqrt{3}-\ln (1+2/\sqrt{3})+\ln (4/3)\simeq 0.67$, which is in fact
not completely negligible compared to $\ln \Delta N_*$ ($\simeq 3.7$ for $\Delta
N_*\simeq 40$).} In general, a classification into universality classes
at next-to-leading order can therefore not be done without largely increasing the number of classes.

\subsection{Validity of the expansion}
\label{sec:validityregime}
In practice, the ``$1/\Delta N_*$ expansion'' is not an expansion in
$1/\Delta N_*$ alone but usually also involves the parameters of the
potential. This implies that the expansion is not always valid, and in
fact, there are potentials for which it is never valid. To illustrate
this statement, let us consider the small field $\sfi_2$ potential
\begin{equation}
  V(\phi) \propto 1-\left(\dfrac{\phi}{\mu} \right)^2,
\end{equation}
which has one free parameter $\mu$. Making use of the techniques
introduced in \Sec{sec:genedef} and defining $x\equiv \phi/\mu$, the slow-roll
trajectory reads
\begin{align}
\xstar=\sqrt{-\Lambert{0}\left(-e^{-\fstar}\right)}\, ,
\end{align}
where $\Lambert{0}$ denotes the ``$0$'' branch of the Lambert
function and $\fstar$ is given by
\begin{equation}
\fstar\equiv 
4\Delta \Nstar \frac{\Mp^2}{\mu^2}+\xend^2(\mu)
-2\ln \left[\xend(\mu)\right].
\label{eq:fstarsfi2}
\end{equation}
In this equation, $\xend$ is the value of $x$ at the end of inflation,
\begin{align}
\xend=\frac{1}{\sqrt{2}}\frac{\Mp}{\mu}
\left(-1+\sqrt{1+\frac{2\mu^2}{\Mp^2}}\right).
\end{align}
In \Eq{eq:fstarsfi2}, $\Delta \Nstar$ appears multiplied by the
dimensionless parameter $\Mp^2/\mu^2$ such that performing an
expansion in $1/\Delta \Nstar$ requires some assumptions on $\mu/\Mp$.
This shows that, as mentioned above, the small parameter of the
expansion is usually not $1/\Delta \Nstar$ alone. In the validity domain 
of the large $\fstar$ limit, the Lambert function can be Taylor expanded according
to $\Lambert{0}(x)\sim x$ and one obtains
\begin{eqnarray}
\label{eq:epsVsfi2:eps1}
\epsVstar{1} & = &\dfrac{\Mp^4}{\mu^4}  \left(\sqrt{1+ 2 \dfrac{\mu^2}{\Mp^2}}-1\right)^2 \nonumber \\ & \times &
\exp\left[-\dfrac{\Mp^2}{\mu^2} \left(4 \Delta \Nstar + 1 + \dfrac{\mu^2}{\Mp^2} - \sqrt{1 +2
    \dfrac{\mu^2}{\Mp^2}}\right)\right] \nonumber \\ & &
   + \order{e^{-8\frac{\Mp^2}{\mu^2}\Delta N_*}}\, , \\
\epsVstar{2} & = &  4 \frac{\Mp^2}{\mu^2} + \order{e^{-4\frac{\Mp^2}{\mu^2}\Delta N_*}}\, .
\label{eq:epsVsfi2:eps2}
\end{eqnarray}
Let us notice that $\epsstar{1}$ does not behave as in
\Eq{eq:expeps1} but belongs to the ``non-perturbative class'' of
Ref.~\cite{Garcia-Bellido:2014gna}.

This expansion is valid as soon as $\fstar \gg 1$, which is the case
if $\mu \ll \Mp$. This limit is, however, inconsistent with the
slow-roll approximation because $\epsVstar{2} \gg 1$. Let us stress
that, as soon as slow roll is violated, one can no longer make use of
\Eqs{eq:ns} and \eqref{eq:r} to derive analytical expressions for the
spectral index and the tensor-to-scalar ratio in terms of the
$\epsstar{n}$, and, thus, in terms of $\Delta \Nstar$. From
\Eq{eq:epsVsfi2:eps2}, one has $\epsVstar{2}<1$ for $\mu> 2 \Mp$. As a
result, \Eq{eq:fstarsfi2} shows that $\fstar$ could be made reasonably
large in the large $\Delta \Nstar$ limit and provided $\mu/(2 \Mp)$ is
of order unity. Only in this very contrived situation
\Eqs{eq:epsVsfi2:eps1} and~(\ref{eq:epsVsfi2:eps2}) might be used.

We conclude that, in general, a $1/\Delta\Nstar$ expansion cannot be
performed for arbitrary values of the free parameters of the model. In
this sense, it is not universal.

\section{Insufficiently accurate}
\label{sec:notsufficient}
In this section, we investigate whether the $1/\Delta N_*$ expansion
of the Hubble-flow parameters is sufficient to match the accuracy of
the present and future data. We choose to exemplify the question with
two models. One is $\si$, the Starobinsky
model~\cite{Starobinsky:1980te} already introduced in
\Sec{sec:noeqClassNLO}. The other one is the small field model
$\sfi_4$, with $V(\phi) \propto 1 - (\phi/\mu)^4$, which has one free
parameter $\mu$. Both models are compatible with the current
data. Moreover, there are values of $\mu$ for which these two models
could a priori be confused, and their disambiguation is a relevant
question for future CMB experiments. The expressions of $\nS$ and $r$
for $\si$ at leading order in slow roll have already been established
in the last section, namely
\begin{equation}
\begin{aligned}
\nS & =  1 - \dfrac{2}{\Delta \Nstar} 
+ \order{\dfrac{\ln \Delta N_*}{{\Delta \Nstar}^2}} ,\\
r & =  \dfrac{12}{{\Delta \Nstar}^2} 
+ \order{\dfrac{\ln \Delta N_*}{{\Delta \Nstar}^2}} .
\end{aligned}
\label{eq:hi:lo}
\end{equation}
For $\sfi_4$, the field trajectory can be solved in terms of $\xstar
\equiv \phistar /\mu$ as
\begin{equation}
\label{eq:xstar:sfi4}
\xstar = \sqrt{\dfrac{\fstar - \sqrt{\fstar^2 -4}}{2}}\, ,
\end{equation}
where $\fstar$ is defined by
\begin{equation}
\label{eq:sfi4:fstar}
\fstar = 8 \Delta \Nstar \dfrac{\Mp^2}{\mu^2} 
+ \xend^2(\mu) + \dfrac{1}{\xend^2(\mu)}\,.
\end{equation}
In the large $\Delta \Nstar \Mp^2/\mu^2$ limit, one has $\xstar\simeq
\fstar/\sqrt{2}\simeq 8\Delta \Nstar \Mp^2/(\sqrt{2}\mu^2)$, which
gives rise to
\begin{equation}
\begin{aligned}
\nS & =  1 - \dfrac{3}{\Delta \Nstar} 
+ \order{\dfrac{1}{{\Delta \Nstar}^2}},\\
r & =  \dfrac{\mu^4}{4 \Mp^4 {\Delta \Nstar}^3} 
+ \order{\dfrac{1}{{\Delta \Nstar}^4}}.
\end{aligned}
\label{eq:sfi4:lo}
\end{equation}
As opposed to $\sfi_2$ discussed in \Sec{sec:validityregime}, one can check that there is no slow-roll
violation for any reasonable values of $\mu$ and the expansion is under control.

\begin{figure}
\begin{center}
\includegraphics[width=0.50\textwidth]{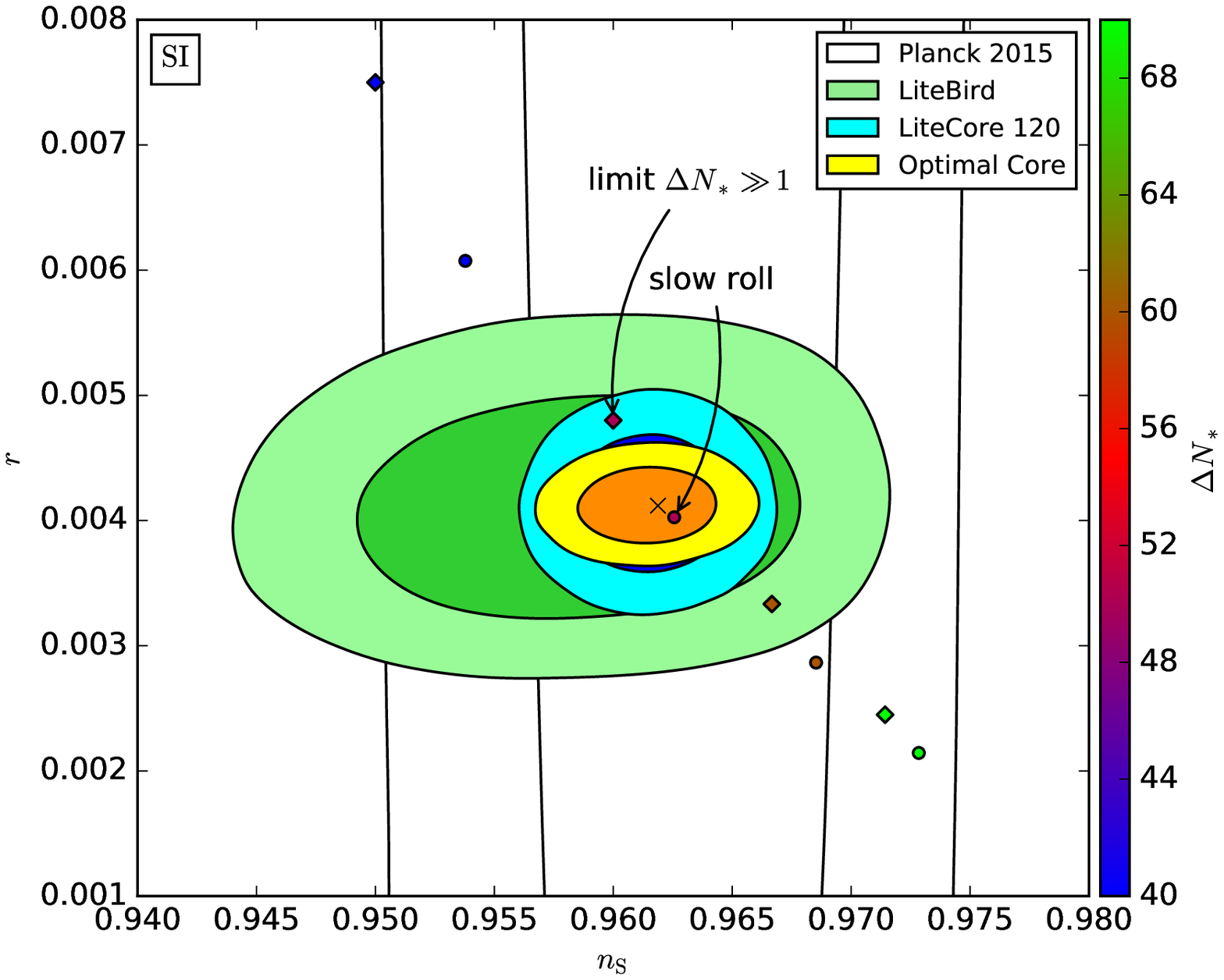}
\includegraphics[width=0.50\textwidth]{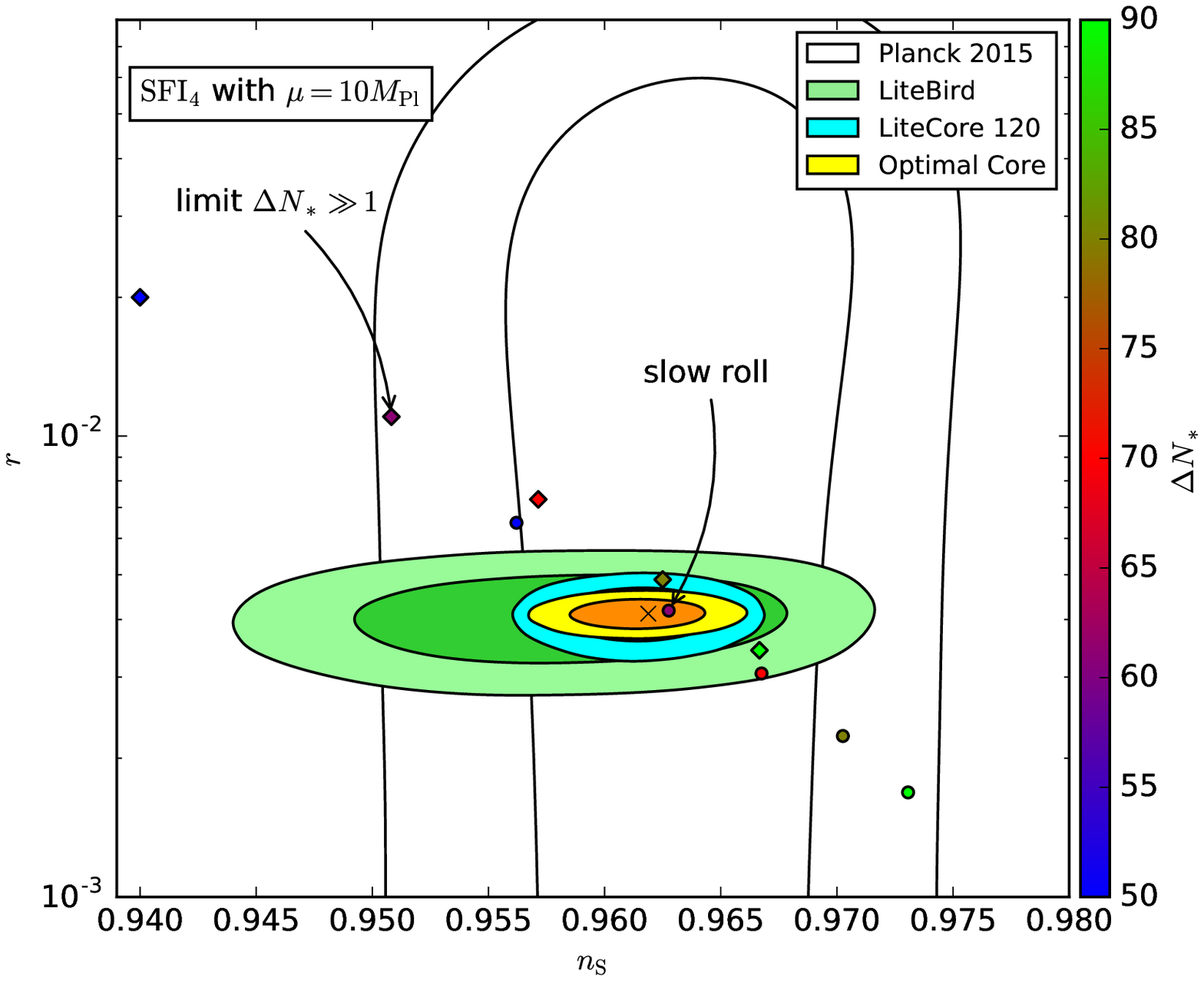}
\caption{Slow-roll predictions (circles) versus $1/\Delta \Nstar$
  expansions (diamonds) in the plane $(\nS,r)$ for various values of
  $\Delta \Nstar$ (color bar). The upper panel shows the expected
  values for Starobinsky inflation ($\si$) while the bottom panel is
  for small field inflation $\sfi_4$ with $\mu=10\Mp$. In both frames, we have
  represented the current one- and two-sigma confidence intervals from
  the Planck 2015 data together with the forecasts of some future
  experiments such as LiteBird and LiteCore (the fiducial model is
  denoted with the black cross). For a fixed value of $\Delta \Nstar$, the
  $1/\Delta \Nstar$ expansion is not sufficiently accurate (the arrows point to the case $\Delta \Nstar = 50$ for $\si$ and $\Delta \Nstar = 60$ for $\sfi4$).}
\label{fig:nsrlo}
\end{center}
\end{figure}

In Fig.~\ref{fig:nsrlo}, for various values of $\Delta \Nstar$
(represented in the color bar), we have plotted as diamonds the
leading order expressions of $\nS$ and $r$ for both $\si$,
\Eq{eq:hi:lo}, and $\sfi_4$, \Eq{eq:sfi4:lo}, together with the
non-approximated slow-roll predictions (circles). For a given value of
$\Delta \Nstar$, the $1/\Delta \Nstar$ expansion significantly
deviates from the non-approximated result.  For Starobinsky inflation
(see upper panel), the expansion in $1/\Delta \Nstar$ yields an
inaccuracy of about half to one sigma compared to the Planck 2015
constraints in the plane $(\nS,r)$. This is barely enough for
assessing the viability of the model with the Planck data. In the same
figure, we have represented the expected constraints of some future
CMB experiments, assuming $\si$ as a fiducial. They are
LiteBird~\cite{Matsumura:2013aja} and various possible designs of the
LiteCore mission~\cite{core}. For ``Optimal Core'', the case perfectly
compatible with the data would appear disfavored by more than three
sigmas if one would trust the result of the $1/\Delta \Nstar$
expansion. The bottom panel of \Fig{fig:nsrlo} shows $\sfi_4$ with
$\mu=10\Mp$. For such a value of $\mu$, and $\Delta \Nstar \simeq 60$,
$\sfi_4$ matches $\si$. However, if one uses the $1/\Delta \Nstar$
expansion, the resulting values of $\nS$ and $r$ are two-sigma away
from the correct values already with the Planck 2015 data, and
completely off with an experiment like LiteCore.

\Fig{fig:nsrlo} suggests that the main source of error in the plane
$(\nS,r)$ is a shift in the value of $\Delta \Nstar$. For $\sfi_4$,
one would need to subtract typically $20$ to $\Delta \Nstar$ to
recover an acceptable result. Let us notice that the uncertainties
associated with slow-roll violations toward the end of inflation may
also induce a discrepancy on $\Delta \Nstar$, but not more than $\pm
1$, which is negligible compared to the effect discussed here. The
issue comes from the expansion itself and the underlying assumption of
considering the large $\Delta \Nstar$ limit.

Let us also mention that we have been fair in choosing the models
displayed in \Fig{fig:nsrlo}, since other models exhibit much larger
departures (as for instance $\sfi_4$ with a larger value of $\mu$).
\section{Incompatible with reheating}
\label{sec:reheat}
Expanding the Hubble-flow parameters in terms of $1/\Delta \Nstar$
raises the question of specifying the value of $\Delta \Nstar$. The
standard lore is to take the values in the range $\Delta \Nstar\in
[50,60]$, or $[40, 70]$, or simply postulate a fixed number like
$\Delta \Nstar=60$. These values may indeed be reasonable but under
various conditions.

One has to make some assumptions on how the reheating era proceeded,
and on the energy scale at the end of inflation. Within a given
inflationary scenario, in which the potential is specified, the energy
scale at which inflation ends is fixed by the model parameters and
this is how the above-quoted numbers can actually be obtained. But
this is no longer the case when one is only interested in expanding
quantities around the pivot scale, as this is done in any of the
$1/\Delta \Nstar$ expansions. For instance, there are inflationary
models without scalar fields in which the Hubble parameter grows
during inflation~\cite{Jimenez:2015jqa} and for which typical values
of $\Delta \Nstar$ could be completely different than $[50,60]$. Even
for single-field inflation, depending on how reheating proceeds,
\Ref{Liddle:2003as} has shown that $\Delta \Nstar = 100$ is
possible. The use of an expansion in $1/\Delta \Nstar$ is therefore
questionable if one cannot predict the value of $\Delta \Nstar$, and
it is easy to check in \Fig{fig:nsrlo} the consequences of taking
$\Delta \Nstar = 40$ or $\Delta \Nstar = 100$ on the predicted values
of $\nS$ and $r$.

The solution to this issue is to specify the inflationary
potential. In this case, $\Delta \Nstar$ is given
by~\cite{Martin:2006rs, Martin:2010kz, Ringeval:2013hfa}
\begin{align}
\label{eq:dnstar}
\Delta \Nstar& =\frac{1-3 \wrehbar}
{12\left(1+ \wrehbar \right)}
\ln \left(\dfrac{\rhoreh}{\rhoend}\right) - \Nzero
\nonumber \\ &
-\dfrac{1}{4} \ln \left[\dfrac{3}{\epsstar{1}}\dfrac{3- \epsstar{1}}{3-\epsoneend}
\dfrac{\Vend}{\Vstar}\right]
+\frac{1}{4} \ln \left(\dfrac{\Hstar^2}{\Mp^2 \epsstar{1}} \right),
\end{align}
where $\Nzero \equiv \ln\left[(\kP /a_0 ) \rhotildegamma \right]$,
$\kP$ being the pivot scale and $\rhotildegamma = \rdofreh
\rhogamma$ with $\rhogamma$ the total energy density stored in
radiation today and $\rdofreh$ a measure of the change of relativistic
degrees of freedom between the reheating epoch and today. Of course,
$V$ denotes the inflationary potential and $\Hstar^2 / \epsstar{1} = 8
\pi^2 \Mp^2 \Pstar + \order{\epsstar{1}}$ where $\Pstar$ is the amplitude of
the scalar power spectrum at the pivot scale. The quantity $\rhoend$
denotes the energy density at the end of inflation (and depends on the
model of inflation) while $\rhoreh$ is the energy density at the end
of reheating. Finally, $\wrehbar$ is the mean equation of state during
reheating. Equation~\eqref{eq:dnstar} shows that once the inflationary
model and the parameters describing reheating are chosen (as well
as the post-inflationary cosmic evolution), $\Delta \Nstar$ is fully
determined. In practice, however, if the inflationary Lagrangian does
not specify the couplings between the inflaton and other sectors, the
reheating parameters are only bounded to vary within specific ranges:
$\rhoreh$ must be larger than the energy density at Big-Bang
Nucleosynthesis (BBN) and smaller than $\rhoend$ while
$-1/3<\wrehbar<1$. This means that there is a completely determined
prior range in which $\Delta \Nstar$ can vary. Postulating fixed values for
$\Delta \Nstar$ misses this fact and 
can lead to incorrect results.

For instance, let us consider the same small field inflationary model
as in \Sec{sec:notsufficient}, namely $\sfi_4$ where we fix $\mu = 10
\Mp$. Moreover, let us assume that reheating has a mean equation of
state given by $\wrehbar=-0.3$, which could, for instance, signal a
low decay rate of the inflaton or the persistence of some vacuum
energy during reheating. An analysis based on \Eq{eq:dnstar} shows
that, for this model, $\Delta \Nstar\in [18.7,55.8]$, the lower bound
being obtained for a reheating at BBN ($\rhonuc^{1/4} = 10\,\MeV$)
while the upper bound corresponds to an instantaneous (or
radiation-like) reheating. Within slow roll, one obtains that the
spectral index $\nS \in [0.904,0.960]$ showing that for this scenario
to be within the two-sigma confidence intervals of the Planck data,
reheating should be almost instantaneous (see \Fig{fig:nsrlo}).

Within the $1/\Delta \Nstar$ expansion formalism, assuming $\Delta
\Nstar \in [40,70]$ would therefore miss most of the physical range of
values while encompassing all the unphysical ones $\Delta \Nstar >
55.8$. For these, the reheating would end at an energy scale higher
than the energy at the end of inflation! Let us stress that this issue
has nothing to do with the inaccuracy of the expansion discussed in
\Sec{sec:notsufficient} and simply comes from the fact that one cannot
arbitrarily choose a fixed number for $\Delta \Nstar$. Nevertheless,
one should notice that the inaccuracy of the expansion makes the
problem even worse. As can be checked in \Fig{fig:nsrlo}, if one uses
the $1/\Delta \Nstar$ expansion and tries to infer the ``right value''
of $\Delta \Nstar$ to make $\sfi_4$ compatible with the Planck data,
one would obtain $\left. \Delta \Nstar \right|_{\nS=0.96} \simeq 75$.

\begin{figure}
\begin{center}
\includegraphics[width=0.50\textwidth]{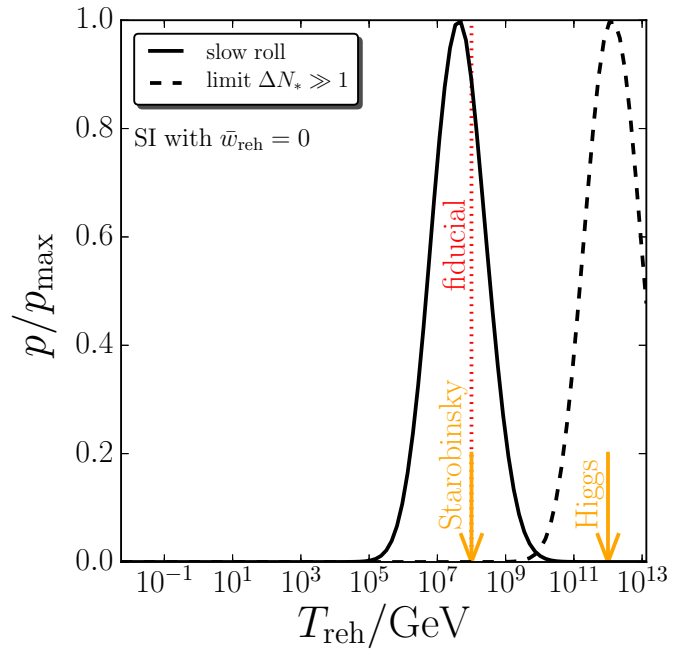}
\caption{Forecast of the marginalized posterior probability
  distribution for the reheating temperature $\Treh$ expected by a CMB
  satellite design such as ``Optimal Core'' (solid curve). The
  fiducial inflationary model is Starobinsky inflation with
  $\wrehbar=0$ and $\Treh=10^8\,\GeV$. The reheating temperature can
  be accurately inferred. The dashed curve shows what would be
  obtained by using the $1/\Delta \Nstar$ predictions. The preferred
  value of $\Treh$ is off by more than five sigmas and would favor a reheating
  scenario typical of Higgs inflation rather than Starobinsky
  inflation, an unfortunate conclusion indeed.}
\label{fig:hireh}
\end{center}
\end{figure}

An incorrect argument against the above discussion would be to postulate
that nothing can be said about the reheating era. As shown in
\Refs{Martin:2010kz, Martin:2014nya, Dai:2014jja, Drewes:2015coa,
  Martin:2016oyk}, the recently released Planck data already allow us
to infer some reheating physics from CMB data. As a result, and even
if the above-mentioned limitations of the $1/\Delta \Nstar$
expansion could be alleviated, one would still miss the opportunity to
constrain reheating.

As an illustration of what the future CMB measurements could tell us
about reheating, we have plotted in \Fig{fig:hireh} the marginalized
posterior distribution of the reheating temperature that can be
inferred by the Optimal Core satellite design (solid curve). Here, the
fiducial reheating history has been assumed to be with a vanishing
equation of state $\wrehbar=0$ and $\Treh = 10^8\,\GeV$, a low value
typical of the reheating after Starobinsky
inflation~\cite{Terada:2014uia}. Not considering the reheating effects
would simply prevent us from making such a measurement.

The dashed curve shows the posterior that would be obtained by using
the $1/\Delta \Nstar$ expansion on $\nS$ and $r$ for extracting the
reheating temperature with \Eq{eq:dnstar}. Let us notice that it would
not make much sense to do so as one would still need the field
potential in this equation.  In any case, the inferred value of
$\Treh$ derived in such a way is off by more than five sigmas and would
wrongly prefer higher reheating temperatures, which are typical of
Higgs inflation~\cite{Bezrukov:2007ep, GarciaBellido:2008ab,
  Figueroa:2015rqa, Repond:2016sol}. As such, using the $1/\Delta
\Nstar$ predictions, one would wrongly conclude that inflation is more
likely to be Higgs rather than Starobinsky.

Reheating is therefore a crucial part of the inflationary
scenario that is now observationally constrained~\cite{Martin:2014rqa}, 
but which cannot be reconstructed with phenomenological expansions. 
\section{Equation-of-state inflation?}
\label{sec:eosi}
All of the previously discussed problems of the $1/\Delta \Nstar$
expansion can be alleviated by simply not performing an expansion at
all. Instead, following Ref.~\cite{Mukhanov:2013tua}, one may decide
to parametrize the inflationary background by specifying the evolution
of the equation of state of the Universe $w(N) = P/\rho$ with respect
to the number of e-folds $N$. This approach was already employed in
\Refs{Barrow:1990td, Barrow:1993zq} and extended in
\Refs{Parsons:1995ew, Barrow:2007zr} by postulating the evolution of
the scale factor $a(t)$ with respect to cosmic time. As shown in these
references, the functional forms chosen for $w(t)$ and $a(t)$ are
equivalent to specifying the inflationary potentials of the so-called
``Intermediate'' and ``Logamediate'' models (II and LMI, see Secs.~5.2
and 5.4 of \Ref{Martin:2014vha}). Here as well, we show that the
choice of $w(N)$ made in \Ref{Mukhanov:2013tua} is equivalent to
choosing a two-parameter potential that we derive.
\subsection{Background evolution}
The hydrodynamical Friedmann-Lema\^itre equations read
\begin{equation}
\begin{aligned}
  H^2 & = \dfrac{\rho(N)}{3 \Mp^2}\, , \quad\quad\quad
\dfrac{\dd H}{\dd N} = -\dfrac{3}{2}\left[1+w(N) \right]H\,.
\end{aligned}
\label{eq:encons}
\end{equation}
As a result, specifying $w(N)$ fixes almost all of the non-perturbed
quantities, up to the integration constant of Eq.~(\ref{eq:encons});
i.e. the energy scale of inflation remains, within this
representation, a fundamentally unpredictable quantity. This is the
first drawback of the hydrodynamical approach. When one specifies an
inflationary potential, the energy scale is fixed by the overall
multiplicative constant, usually referred to as $M^4$. For most of the
inflationary models proposed so far, this parameter is usually not
predicted by the theory and chosen to match the amplitude of the CMB
anisotropies. In that situation, specifying $w(N)$ is indeed not worse
than letting $M^4$ be a free parameter. However, there are
inflationary models for which $M^4$ is predicted. For instance, this
is the case for the very first models of inflation such as Starobinsky
Inflation ($\si$)~\cite{Starobinsky:1980te}, Higgs
inflation~\cite{Bezrukov:2007ep}, the original Coleman-Weinberg model
($\cwi$)~\cite{Guth:1980zm, Linde:1981mu} (ruled out for this very
reason~\cite{Abbott:1981rg, Ellis:1982ws, Albrecht:1983ib}), Open
String Tachyon Inflation
($\osti$)~\cite{Witten:1992qy,Witten:1992cr,Gerasimov:2000zp,Kutasov:2000qp}
(also ruled out for this reason~\cite{Kofman:2002rh}) and Dual
Inflation ($\di$)~\cite{AlvarezGaume:1996gd,
  GarciaBellido:1997mq}. Compared to these, an inflationary background
evolution given by $w(N)$ remains less predictive.

More interestingly, one can rewrite the Friedmann-Lema\^itre equations
in terms of the first Hubble-flow function $\eps{1}$. From its
definition, one gets
\begin{equation}
  \eps{1}(N) \equiv - \dfrac{\ud \ln H}{\ud N} = \dfrac{3}{2} \left[1 + w(N) \right].
\label{eq:eps1w}
\end{equation}
As a result, specifying the equation-of-state is \emph{strictly}
equivalent to postulating the evolution of the first Hubble-flow
function $\eps{1}(N)$. The complete Hubble-flow hierarchy
$\eps{n}(N)$ is then exactly known. For instance, the second and third
Hubble-flow functions read:
\begin{equation}
\eps{2}(N) = \dfrac{\dot{w}(N)}{1 + w(N)}  \,, \quad
\eps{3}(N) = \dfrac{\ddot{w}(N)}{\dot{w}(N)} - \dfrac{\dot{w}(N)}{1 + w(N)}\,,
\end{equation}
where a dot denotes here the derivative with respect to
$N$. \Eq{eq:eps1w} also determines $\Nend$, the e-fold at which
inflation ends, given by solving $\eps{1}(\Nend)=1$. As a result,
$\Delta \Nstar=\Nend - \Nstar$ is well defined and, up to the unknown
integration constant of Eq.~(\ref{eq:encons}), the energy scale at
which inflation ends can be uniquely determined. In particular, this
allows the reheating era to be consistently considered and $\Delta
\Nstar$ to be determined. At this point, one may wonder what the
difference is, at the background level, compared to the more usual
situation in which one specifies the field potential. The answer is
none.

Indeed, comparing the Friedmann-Lema\^itre equations obtained from a
minimally coupled scalar field to the hydrodynamical
ones~\cite{Martin:2013tda}, one gets
\begin{equation}
\begin{aligned}
\left(\dfrac{\ud \phi}{\ud N}\right)^2 & = 2 \Mp^2 \eps{1}(N),\\
\dfrac{\ud \ln V(\phi)}{\ud N} & = -2 \eps{1}(N) + \dfrac{\ud \ln
  [3-\eps{1}(N)]}{\ud N}\,.
\end{aligned}
\label{eq:kgw}
\end{equation}
Using \Eq{eq:eps1w}, these equations can be formally integrated
as
\begin{equation}
\begin{aligned}
\phi(N) & = \phizero \pm \sqrt{3} \, \Mp  \int_{N_0}^N \sqrt{1 + w(n)} \, \ud n,\\
V(N) &= \Vzero \frac{1- w(N)}{1- w(N_0)} \exp\left\{-3 \int_{N_0}^N \left[1+w(n) \right] \ud n\right\},
\end{aligned}
\label{eq:Vw}
\end{equation}
where $\Vzero$ is the expected integration constant associated with
energy conservation. The other integration constant, $\phizero$, has
no observable effect and comes from the shift symmetry of
\Eq{eq:kgw}, while $w(N_0)$ can be absorbed in $V_0$. 
\Eq{eq:Vw} gives a parametric
representation of the field trajectory and its potential. Solving for
$\phi(N)$, one then infers $N(\phi)$ which leads to
$V(\phi)=V[N(\phi)]$.

As an illustration, let us recover the exact field potential
associated with
\begin{equation}
\label{eq:eps1star}
1+w \equiv \dfrac{\beta}{\left(c + \Delta \Nstar\right)^{\alpha}}\,,
\end{equation}
where $\alpha$ and $\beta$ are two free parameters and, following
Ref.~\cite{Mukhanov:2013tua}, $c$ is a regularizing constant to avoid
any divergences at the end of inflation. Let us stress that the above
equation is a definition and not an expansion as in
\Eq{eq:expeps1}. In order to consistently implement the end of
inflation $\eps{1}(\Nend)=1$ [or, equivalently, $w(\Nend)=-1/3$], one
has to fix $c = (3 \beta/2)^{1/\alpha}$. Integrating Eq.~\eqref{eq:Vw}
and fixing $\phizero=\mp\sqrt{3 \beta}/(1-\alpha/2)$ gives the
potential (some approximations of which are obtained in \Ref{Mukhanov:2013tua})
\begin{equation}
\begin{aligned}
V(\phi) & = M^4 \left[1 - \dfrac{\beta}{2
    \left(1+\dfrac{2-\alpha}{2\sqrt{3\beta}}\dfrac{\phi}{\Mp}\right)^{\frac{2\alpha}{2-\alpha}}}
  \right] \\ & \times  \exp\left\{ \dfrac{3 \beta}{1-\alpha}
\left[\left(1+\dfrac{2-\alpha}{2\sqrt{3\beta}}\dfrac{\phi}{\Mp}\right)^{\frac{2(1-\alpha)}{2-\alpha}}
  - 1 \right]
\vphantom{\dfrac{\beta}{2
    \left(1+\dfrac{2-\alpha}{2\sqrt{3\beta}}\dfrac{\phi}{\Mp}\right)^{\frac{2\alpha}{2-\alpha}}}}
\right\}.
\end{aligned}
\label{eq:vfmipot}
\end{equation}

In the limit $\alpha \rightarrow 1$, it is interesting to notice that
this is nothing but the potential of Intermediate Inflation, see
Sec.~5.2 of \Ref{Martin:2014vha}. We also recover explicitly the
result of Ref.~\cite{Mukhanov:2013tua}: for $\alpha < 1$, the
potential has an exponential shape reminiscent of Power Law Inflation
($\pli$) and Logamediate Inflation ($\lmi$) (although $\lmi$ is
defined with a relation among the coefficients characterizing the
potential which is not obtained in the present case), for $1<\alpha
\le 2$ it is of the plateau kind and for $\alpha > 2$ it is similar to
Small Field Inflation [$\sfi_{p}$ with $p=2\alpha/(\alpha-2)$].

We conclude that, for the background evolution, choosing a function
$w(N)$, or equivalently $\epsilon_1(N)$, is not a generic procedure
but just singles out a
particular $V(\phi)$, namely a particular model of inflation. The only
difference with respect to the traditional approach is that the energy
scale of inflation can no longer be predicted.

\subsection{Cosmological perturbations}

Specifying $w(N)$ instead of $V(\phi)$ is, however, not enough to
uniquely determine the behavior of the cosmological perturbations
during inflation~\cite{Malik:2004tf, Chen:2013kta}. Indeed, if
$\bardeen$ represents (the Fourier transform of) the Bardeen potential
and if the Universe is dominated by a perfect fluid, then one has
\begin{equation}
\begin{aligned}
\bardeen''& +3{\calH}\left(1+\cS^2\right)\bardeen'
+\left[2{\calH}'+{\calH}^2 \left(1+3 \cS^2\right)\right]\bardeen \\ &
+\cS^2 k^2 \bardeen = \frac{a^2}{2\Mp^2} \delta \pnad,
\end{aligned}
\label{eq:bardeen}
\end{equation}
where a prime denotes a derivative with respect to conformal time and
$\calH = a H$ is the conformal Hubble parameter. In this expression
$\cS^2 \equiv P'/\rho'$ is the sound speed and $\delta \pnad \equiv
\delta P - \cS^2 \delta\rho $ is the non-adiabatic pressure
perturbation. If one wants the hydrodynamical perturbations to evolve
as the perturbations stemming from a scalar field, the sound speed
must verify the relation
\begin{equation}
\begin{aligned}
\cS^2 = 1 & - \dfrac{4}{9[1-w(N)^2]} 
\left\{3+3 w(N) - \dfrac{\ud \ln [1-w(N)]}{\ud N} \right\},
\end{aligned}
\label{eq:soundspeed}
\end{equation}
while the fluid must possess a non-adiabatic pressure such that
\begin{equation}
\label{eq:pnad}
\delta \pnad = -2 \Mp^2 \left(1 - \cS^2\right) \dfrac{k^2}{a^2} \bardeen.
\end{equation}
From \Eqs{eq:eps1w} and \eqref{eq:kgw}, one indeed recovers the speed of sound
associated with a perturbed scalar field
\begin{equation}
\cS^2 = 1 + \dfrac{2 a^2 V_{,\phi}}{3 \calH \phi'}\,.
\end{equation}
Inserting \Eqs{eq:soundspeed} and~\eqref{eq:pnad} in
\Eq{eq:bardeen} leads to an equation for the Bardeen potential
which is exactly that obtained under the assumption that the dominant
fluid in the Universe is a scalar field~\cite{Mukhanov:1990me}.

As a result, and as opposed to a scalar field, it is not sufficient to
specify the background, namely the function $w(N)$, to fix the
evolution of the perturbed quantities. One should also specify the
functional form of $\cS^2(N)$ and $\delta \pnad(N)$ to have
well defined equations of motion. Conversely, implicitly assuming that
the hydrodynamical perturbations evolve as the ones generated during
single-field inflation, one must have a very contrived sound speed
$\cS^2(N)$ and non-adiabatic pressure $\delta \pnad(N)$. It is hard to
understand how this could be achieved without the knowledge of
Eqs.~\eqref{eq:soundspeed} and~\eqref{eq:pnad}, namely without knowing
that the underlying model is, as a matter of fact, a scalar field.

\section{Statistically flawed}
\label{sec:stat}
Finally, let us discuss whether phenomenological parametrizations of inflation are well suited to carry out statistical model comparison.

We consider a model of inflation characterized by the parameters
$\thetai$ [including the mass scale $M$ of the potential and any other
  parameters needed to completely specify the shape of $V(\phi)$ such
  as $\mu$ for $\sfi$] and $\thetar$ (the reheating parameters, see
\Sec{sec:reheat}). In the slow-roll approximation, the power
spectra of tensor and scalar perturbations are functions of the
Hubble-flow parameters $\epsstar{n}$ only, which, in turn, are
functions of the $\thetai$ and $\thetar$ parameters. As a consequence, the
predictions of a model in terms of the primordial power spectra are
expressed with ${\calP}_{\zeta}(\thetai,\thetar)$ and
${\calP}_{h}(\thetai,\thetar)$. In this manner, the slow-roll
approximation is a proxy to facilitate the derivation of the power
spectra as functional of the underlying theory parameters, exactly as
one would obtain from an exact integration of the inflationary
perturbations~\cite{Ringeval:2013lea, Ringeval:2007am}. This is a
crucial difference between slow roll and the previously discussed
alternatives which discard any underlying theoretical model.

However, in order to estimate the statistical ability of a hypothesis
to explain the observed data~\cite{Trotta:2008bp, Trotta:2008qt,
  Ringeval:2013lea, Martin:2013nzq, Martin:2014lra, Price:2015qqb},
one must first specify the prior distributions of the underlying
parameters. For the inflationary models, they are the $\thetai$ and
$\thetar$ parameters and their prior distribution naturally stems from the
underlying theoretical assumptions.

Instead, starting only with, say, $\epsstar{1}=\beta \left(\Delta
\Nstar\right)^{-\alpha}$, there is no guidance to choose the priors on
$\alpha$, $\beta$ (for the inconsistencies in choosing $\Delta
\Nstar$, see \Sec{sec:reheat}). In the absence of any other
information, a simple guess would be, for instance, to take a flat
prior on $\alpha$ and $\beta$. But if the purpose of $\epsstar{1}(N)$
is to actually represent an inflationary model, then
$\alpha=\alpha(\thetai)$ and $\beta =\beta (\thetai)$ such that flat
priors on $\alpha$ and $\beta$ would correspond to unnatural 
priors on the $\thetai$ and $\thetar$ parameters. As a matter of
fact, Bayesian evidence derived in such a way would be flawed.

Let us now illustrate the above considerations with a very simple
model, Loop Inflation ($\li$), the potential of which is given by
$V(\phi)=M^4\left[1+\alphali \ln(\phi/\Mp)\right]$. It is
characterized by two parameters, the mass scale $M$ and $\alphali$. We
therefore have $\thetai=\lbrace M,\alphali \rbrace$. At leading order
in slow roll, $\li$ is a model like $\sfi_2$ for which the expansion
in $1/\Delta \Nstar$ does not enter any known classification. An
expansion in $\alphali$ may, however, be consistently performed and one
gets
\begin{equation}
\begin{aligned}
\epsstar{1} &= \frac{\alphali}{\Delta \Nstar} +
\orderb{\alphali^2\ln(\alphali \Delta\Nstar)}, \\
\epsstar{2} &= \frac{1}{\Delta \Nstar} + \order{\dfrac{\alphali}{\Delta \Nstar}}.
\end{aligned}
\label{eq:epsLI}
\end{equation}
This means that, for this model, $\alpha (\thetai)=1$ and $\beta
(\thetai)=\alphali/4$. Ignoring the underlying model and just
postulating $\epsstar{1}=\beta \left(\Delta \Nstar\right)^{-\alpha}$,
one would be tempted to choose a flat prior on $\beta$, i.e. a flat
prior on $\alphali$. But within Loop Inflation, $\alphali$ is a
coupling constant as the logarithm in the expression of the potential
originates from a one-loop calculation. As a consequence $\alphali$ is
a small parameter, the order of magnitude of which is unknown a
priori. Therefore, an uninformative prior for $\alphali$ is a
Jeffreys' prior. Assuming a flat prior would not lead to equal
probability per decade and would bias $\alphali$ toward
unnatural large values.

It is then worth recalling that changing the prior may modify 
the posterior since 
\begin{equation}
  P(\alphali \vert D)=\dfrac{1}{P(D)}{\calL}
(D \vert \alphali)\pi(\alphali).
\end{equation}
Depending on how peaked the likelihood $\calL(D|\alphali)$ is,
different $\pi(\alphali)$ would lead to different $P(\alphali \vert
D)$. More importantly, an incorrect prior would also change the global
likelihood $P(D)$, and thus the Bayesian evidence. As an illustration,
let us consider a toy likelihood function which is a simple Gaussian
\begin{equation}
\calL = \calLmax e^{-\alphali^2/(2\sigma^2)},
\end{equation}
then for a flat prior $\pi^\flat(\alphali)=1/\Delta \alphali$ with $\Delta
\alphali \equiv \alphalimax-\alphalimin$, one obtains
\begin{equation}
\begin{aligned}
P^\flat(D) = \calLmax \sqrt{\dfrac{\pi}{2}}
\dfrac{\sigma}{\Delta \alphali}
\biggl[\erf \left(\frac{\alphalimin}{\sigma \sqrt{2}}\right)
-\erf\left(\frac{\alphalimax}{\sigma \sqrt{2}}\right)\biggr],
\end{aligned}
\end{equation}
where $\erf(x)$ is the error function. For a Jeffreys' prior
$\pi^{\natural}(\alphali)=1/\left[\alphali \ln
  \left(\alphalimax/\alphalimin \right)\right]$, one has
\begin{equation}
\begin{aligned}
P^\natural(D) & = \frac{\calLmax}{2 \ln \left(\alphalimax/\alphalimin
  \right)} \\ & \times \left\{\E_1\left[
  \frac{(\alphalimin)^2}{2\sigma^2}\right] -
\E_1\left[\frac{(\alphalimax)^2}{2\sigma^2}\right] \right\},
\end{aligned}
\end{equation}
where $\E_1(z)=\int_z^{\infty} \ud t e^{-t}/t$ is an exponential
integral function. Viewed as functions of $\alphalimin$ and
$\alphalimax$, the previous ``toy model calculation'' illustrates the
fact that the Bayesian evidence can be very different according to
assumptions made on the prior distributions for the $\thetai$'s.

We therefore conclude that considering $\epsstar{1}=\beta
\left(\Delta \Nstar\right)^{-\alpha}$ without reference to an
underlying theoretical framework leads to uninformative statistical
results. If one ignores the fact that $\alpha =\alpha(\thetai)$ and
$\beta =\beta(\thetai)$, our ability to fix different priors for
different models is lost. As a consequence, this approach is not well suited
to carry out model comparison and derive statistical constraints on the physics
of the early Universe.

\section{Conclusions}
\label{sec:conclusions}
In this short article, we have argued that it is often too simplistic
to view inflation as a framework that can be ``described by two
numbers''. The goal of a model is not to predict the values of $\nS$
and $r$ only. In fact, it should first predict the amplitude of the
cosmological perturbations, as some models actually do ($\si$, $\cwi$,
$\osti$, $\di$). Then, even if inflation is featureless, single field,
slowly rolling, with minimal kinetic terms, one can still reasonably
hope to measure other numbers, such as the running $\alphaS$. But more
importantly, inflation does not only consist in a phase of accelerated
expansion. The mechanism that ends inflation is also of crucial
importance and, as a matter of fact, can be constrained by CMB
data~\cite{Martin:2010kz, Martin:2014nya, Martin:2016oyk}. The new
parametrizations miss this opportunity. They can never be as
informative as an approach rooted in field theory, or some specific
modified gravity framework~\cite{DeFelice:2010aj, DeFelice:2011uc},
when it comes to a phenomenon that could have taken place at an energy
scale as high as $10^{16}\,\GeV$~\cite{Bezrukov:2014ipa}. At last, specifying a model in the
hope of comparing it with some data also means giving the priors on
its free parameters to ensure its internal consistency. This is
usually much more than specifying two numbers.

The price to pay is that some predictions do depend on the underlying
model, but not all of them. For instance, a generic prediction of
inflation is the presence of Doppler peaks in the CMB which makes
inflation a falsifiable scenario. On the other hand, there is no
generic prediction for $r$, except that it must be such that the
energy scale of inflation is higher than the one of BBN, leading to a
ridiculously small lower bound, $r \gtrsim 10^{-75}$, a value which is
unobservable as smaller than backreaction
effects~\cite{Martineau:2007dj}. But this does not necessarily mean
that the situation is not interesting, models do predict different
ranges of tensor-to-scalar ratio values and measuring $r$ provides
information about the underlying inflationary scenario.

One of the goals of phenomenological parametrizations was to narrow
down these ranges and yield ``typical'' inflationary predictions.  For
instance, it is often argued that while
$\epsstar{1}=\order{1}/\Delta\Nstar$ (yielding $r\simeq 0.26$ for
$\Delta\Nstar = 60$) is now excluded by the data, the next target
according to \Eq{eq:expeps1} would be to try and detect the next order
in $1/\Delta\Nstar$, namely $\epsstar{1}=\order{1}/\Delta\Nstar^2$
(yielding $r\simeq 0.004$ for $\Delta\Nstar = 60$). However, nothing
guarantees that the overall constant is indeed of order one. For
instance, as can be seen in \Eq{eq:epsLI}, this is the case for the
model $\li$
since $\alphali \ll 1$. In fact, a value less than $0.25$ is already
sufficient to reestablish the agreement between the prediction
$\epsstar{1}=\order{1}/\Nstar$ and the data.

In conclusion, it seems to us that even if the phenomenological
parametrizations discussed in the present work may provide useful
rule-of thumb classifications, the most promising method to learn
about the physics of inflation is to build models based on high energy
physics and (modified) gravity, since this is a priori the way Nature
has realized inflation in practice.  At the time when the Planck data
tell us that the Higgs field of Particle Physics, some low energy
String compactifications, or the $R^2$ corrections to General
Relativity~\cite{Martin:2013nzq}, could explain the large scale
structure of the Universe, it seems that phenomenological
parametrizations are not sufficient to tackle the physical questions
we now have to address. The fact that some predictions are model
dependent is not a shortcoming but actually a virtue of inflation
since it can be used to learn about Physics in a regime hardly
achievable with current technology.

\acknowledgments
It is a pleasure to thank Diederik Roest for interesting discussions
and useful comments. V.~V.'s work is supported by STFC Grants
No. ST/K00090X/1 and No. ST/L005573/1. C.~R. is partially supported by
the Belgian Federal Office for Science, Technical \& Cultural Affairs
through the Interuniversity Attraction Pole P7/37.

\bibliography{biblio}

\end{document}